\documentclass[10pt]{article}

\usepackage{longtable}
\usepackage{amsmath}
\usepackage{amssymb}
\usepackage{graphics}
\usepackage{rotating}
\usepackage{cite}
\usepackage{color}
\usepackage{fancybox}

\def\0{\mbox{\tiny $0$}}
\def\1{\mbox{\tiny $1$}}
\def\2{\mbox{\tiny $2$}}
\def\3{\mbox{\tiny $3$}}
\def\4{\mbox{\tiny $4$}}
\def\5{\mbox{\tiny $5$}}
\def\6{\mbox{\tiny $6$}}
\def\7{\mbox{\tiny $7$}}
\def\8{\mbox{\tiny $8$}}
\def\9{\mbox{\tiny $9$}}
\def\T{\mbox{\tiny T}}

\title{\shadowbox{\large \bf
LASER INTERACTION WITH A DIELECTRIC BLOCK}}

\author{
\small  Stefano De Leo\thanks{ Department of Applied Mathematics,
State University of Campinas, Brazil [deleo@ime.unicamp.br] } \,\,
and\, Pietro Rotelli\thanks{Department of Physics, University of
Salento and INFN Lecce, Italy [rotelli@le.infn.it]}}

\date{\small
\fcolorbox{black}{yellow} {\color{red} $\bullet$ {\color{black}{
{\footnotesize {\sc European Physical Journal D} {\bf 61} (2011)
481-488}}} {\color{red}{$\bullet$}} } }

\begin{document}
%
\maketitle

\vspace*{-.7cm}

\begin{abstract}
\noindent Some optical experiments provide the easiest way to test
quantum mechanical predictions. Such a situation applies to a
laser beam traversing a dielectric structure. The dielectric
structure mirroring quantum mechanical potentials. The simplest of
these, studied in this paper, involves a single dielectric block.
We exhibit both by analytic and numerical calculations explicit
examples of total coherence and total incoherence phenomena. Which
case appears depends upon the angle of incidence of the laser beam
and/or the dimensions of beam vs block width. Unlike our previous
studies our calculations here are three-dimensional, although,
with suitable approximations, somewhat simpler planar expressions
can be derived.
\end{abstract}
















\section*{\normalsize I. INTRODUCTION}

Everything in nature is quantum mechanical at its core but, as the
history of Physics teaches us, many processes can be studied
through ``classical'' equations. More specifically with equations
that do not involve explicitly (or implicitly as occurs when using
natural units) Planck's constant. Examples are Newton's equation,
thermodynamic equations, classical field equations and amongst
these, of particular interest to us, Maxwell's
equations\cite{Born99}. Maxwell's equations encompass optical
phenomena which have a surprising aspect. Under certain
conditions\cite{Abeles50,OI90}, which we recall in the next
section, the equations of say the electric field is analogous to
the non relativistic Schrodinger equation\cite{Cohen77}.  The
potentials in Quantum Mechanics (QM) are identifiable with
dielectric materials with different refractive indices with which
the laser interacts. One of the advantages of this fact is that QM
effects, probabilistic in nature and valid for massive particle
wave packets, can be reproduced, and the theory tested far easier,
with optical analog
experiments\cite{MSG03,SKP07,Ref1,Ref2,Ref3,Ref4,Ref5,Ref6}.

In this paper we wish to simulate one of the simplest of all QM potentials,
i.e. a square potential. We present both analytic and numerical calculations
of a gaussian laser incident upon a single block dielectric. We shall in
particular concentrate upon the transmitted beam or beams. It will be shown
that in the so-called wave limit, where coherence between individual amplitude
terms dominates, a single outgoing laser beam emerges. In the other limit,
when incoherence dominates, multiple laser beams are transmitted. We shall
calculate both the intensity and position of the first few of these beams. Our
intention is to apply these methods  to more complex systems, such as twin
barrier diffusion and tunnelling\cite{ORJ04,W06}. The basic ingredients of our
approach will be detailed in this paper. This work differs from some previous
papers of the present authors\cite{BDR04,DR05,DR06,DR08} by calculating in
three dimensions and by the explicit use of a laser as the light source. This
provides for a more realistic theoretical study, closer to laboratory
experiments. For some of the analytical calculations we shall make use of the
stationary phase method, SPM\cite{W55,BH75}. We shall demonstrate in detail
how our results depend critically upon the laser angle of incidence with the
dielectric block.

In the next section we derive the gaussian laser amplitude in
three-dimensional free space. In section III we describe the geometry of the
laser with respect to a dielectric block corresponding to a QM well. Also in
this section we derive the modification to the laser beam within the
dielectric. In section IV we derive the transmission and reflection amplitudes
for the laser beam (or beams) exiting the dielectric. The method of
calculation follows the so called two step approach\cite{BDR04,DR05,DR06}.
This yields an infinite number of contributing amplitudes. We study the
conditions for which complete coherence occurs yielding a single outgoing
beam. In the other extreme infinite outgoing beams are predicted. We calculate
the exit points and intensities of the leading contributions as a function of
the incoming angle. In section V we calculate the first few incoherent beam
intensities and their exit points from the dielectric block. This is performed
both analytically and numerically. For this analysis typical helium-neon laser
parameters  are employed\cite{Weber01}.

\section*{\normalsize II. THE GAUSSIAN LASER AMPLITUDE}

 Throughout this paper we consider a fixed frequency
 $\omega=|\boldsymbol{k}|/c$. A fixed frequency implies a
 stationary system, so we can forego writing the time dependence
 $e^{-i\omega t}$. For the theoretical analysis we start with plane waves.
 These correspond to momentum eigenvalues and determine a fixed direction
 $\boldsymbol{k}$. A general solution can then be obtained by convoluting
plane waves with a suitable integral.

Consider a plane wave moving in the $\boldsymbol{k}$-direction, the electric
field vector can be then  expressed by
\[\boldsymbol{E}=E\,\boldsymbol{\hat{s}} \propto
\exp^{i\,(k_x x+k_y y+k_z z)}\,\boldsymbol{\hat{s}}\,\,,\] where
$\boldsymbol{\hat{s}}$ is a unitary polarization vector, assumed linear, which
lies in the plane perpendicular to $\boldsymbol{k}$. The condition
$\mbox{div}\boldsymbol{E}=0$ imposed by one of the Maxwell equations implies
$\boldsymbol{k}\cdot \boldsymbol{\hat{s}}=0$. This means that
$\boldsymbol{\hat{s}}$ is $\boldsymbol{k}$  dependent. Now a possible but not
unique choice for $\boldsymbol{\hat{s}}$ is
\[
\boldsymbol{\hat{s}} =
\frac{k_z}{k}\,\left(\,\frac{k_x}{\sqrt{k_x^{^{2}}+k_y^{^{2}}}}\,,\,
\frac{k_y}{\sqrt{k_x^{^{2}}+k_y^{^{2}}}}\,,\,-\,
\frac{\sqrt{k_x^{^{2}}+k_y^{^{2}}}}{k_z}\, \right)\,\,.
\]
We claim this is not unique since, as we vary the direction, we may also
rotate the above $\boldsymbol{\hat{s}}$ about the plane wave direction with
impunity. Furthermore, as given, these do not determine a unique
$\boldsymbol{\hat{s}}$ in the limit $k_{x,y}\to 0$. In fact the direction of
$\boldsymbol{\hat{s}}$ depends upon how these limits are performed. If first
we take $k_x\to 0$ and then $k_y\to 0$ we find $\boldsymbol{\hat{s}}
=(0,1,0)$. If we first take $k_y\to 0$ and then $k_x\to 0$ we find
$\boldsymbol{\hat{s}} =(1,0,0)$. A general direction in the $xy$-plane is
obtained by fixing the ratio of $k_x/k_y$ and taking simultaneously
$k_{x,y}\to 0$. All of the above tells us that the components of
$\boldsymbol{E}$ are not the same and thus, after integrations with a chosen
convolution function, they will yield {\em different} component functions of
$\boldsymbol{E}$. However, if use is made, as we shall do, of a peaked
gaussian function that limits all significant values of wave number to
$k_{x,y}\ll k_z\approx k$ we can approximate $\boldsymbol{\hat{s}}$ by a fixed
vector in the $xy$-plane. The calculation of $\boldsymbol{E}$ then greatly
simplifies to the calculation of the amplitude $E$.

In  free space the gaussian laser is obtained by convoluting a plane wave with
a peaked gaussian,
\[ G(k_x,k_y)=\exp\left[\,-\,\frac{\mbox{w}_{\0}^{\2}}{4}\,
\left(\,k_x^{^{2}}+k_{y}^{^{2}}\, \right)\right]\,\,,\]
 where the arguments are the
 components of the wave number in the direction orthogonal to the
 laser direction, here chosen as the $z$ direction, and with
a Dirac delta  function,
\[
\delta(\,k_z-\sqrt{k^{^{2}}-k_x^{^{2}}-k_y^{^{2}}}\,\,)\,\,.
\]
The electric field amplitude, $E(x,y,z)$, is then proportional to
\[
 \int
\mbox{d}^{\3}k\,\,G(k_x,k_y)\,
\delta(\,k_z-\sqrt{k^{^{2}}-k_x^{^{2}}+k_y^{^{2}}}\,\,)
\,\exp\left[\,i\,(k_x\,x+k_{y}\,y+ k_z\,z\,)\, \right]\,\,.
\]
The integral over $\mbox{d}k_z$ is straightforward because of the Dirac delta
function. So we have
\begin{equation}
\label{las1}
 E(x,y,z) =
E_{\0}\,\frac{\mbox{w}^{\2}_{\0}}{4\,\pi}\,\, \int
\mbox{d}k_x\,\mbox{d}k_y\,\,G(k_x,k_y)\,\exp\left[\,i\,\left(k_x\,x+k_{y}\,y+k_z\,z\,\right)\,
\right]\,\,,
\end{equation}
where $E_{\0}=E(0,0,0)$.  The use of a sharp gaussian also allows us to
approximate the electric field amplitude above as follows
\begin{equation}
\label{l1}
 E(x,y,z)\approx
E_{\0}\,\frac{\mbox{w}^{\2}_{\0}}{4\,\pi}\,\,e^{ikz}\, \int
\mbox{d}k_x\,\mbox{d}k_y\,\,G(k_x,k_y)\,\exp\left[\,i\,\left(k_x\,x+k_{y}\,y-
\frac{k_x^{^{2}}+ k_y^{^{2}}}{2k}\,z\,\right)\, \right]\,\,.
\end{equation}
 From this, after performing two  generalized gaussian
integrations, we obtain the well known closed formula which describes the free
propagation of a gaussian laser beam\cite{EP04}, i.e.
\begin{equation}
\label{l2}
 E(x,y,z)
  \approx  E_{\0}\,\exp[\,i\,k\,z\,]\,\,
\frac{\mbox{w}_{\0}^{\2}}{\mbox{w}_{\0}^{\2}+2\,i\,z/\,k}\, \exp\left[\,-\,
\frac{x^{^{2}}+y^{^{2}}}{\mbox{w}_{\0}^{\2}+2\,i\,z/\,k} \right]\,\,.
\end{equation}
 The intensity distribution of the gaussian beam,
$I(x,y,z)=|E(x,y,x)|^{^{2}}$, is then given by
\begin{equation}
I(x,y,z) \approx  \frac{2\,P}{\pi\,\mbox{w}^{\2}(z)}\,\exp\left[\,-\,2\,\,
\frac{x^{^{2}}+y^{^{2}}}{\mbox{w}^{\2}(z)} \right]\,\,,
\end{equation}
where
\[
\mbox{w}(z)=\mbox{w}_{\0}\,\sqrt{1+\left(\,\frac{\lambda\,\,z}{\pi\,\mbox{w}_{\0}^{\2}}\,
\right)^{^{2}}}
\]
is the radius of the $1/e^{\2}$ contour after the wave has propagated a
distance $z$ and $P$ the total power in the beam,
\[
P =\int \mbox{d}x\,\mbox{d}y\,I(x,y,z) \approx
\frac{\pi\,\mbox{w}_{\0}^{\2}}{2}\,I_{\0}\,\,,
\]
with $I_{\0}=I(0,0,0)$.

In Fig\,1 we plot a side view of the laser intensity for different beam
waists, $\mbox{w}_{\0}=1\,\mbox{mm}$ and $\mbox{w}_{\0}=2\,\mbox{mm}$, and for
a wavelength value of $\lambda=2\pi/k=633\,\mbox{nm}$. These figures display
the fact that the spread in momentum is complementary to the spatial laser
width. Indeed with the angular opening $\theta$ defined as
\[
\tan \theta = \lim_{z\to \infty}
\frac{\mbox{w}(z)}{z}=\frac{\lambda}{\pi\,\mbox{w}_{\0}}\,\,,
\]
the $1/e^{\2}$ intensity contours asymptotically approach a cone of angular
radius $\theta$. For our choice of lambda and for a beam waist of the order of
mm ($\lambda\ll \pi \mbox{w}_{\0}$) the angular radius is then given by
\[ \theta \approx \lambda\,/\,\pi\mbox{w}_{\0}\,\,.\]
This angle is indicative of the spread in wavelength and as given by
$G(k_x,k_y)$ decreases with increasing w$_{\0}$. On the contrary, as seen by
the explicit form of the intensity $I(x,y,z)$, the width of a laser, for any
chosen value of $z$, grows with w$_{\0}$. As an aside we recall that the wave
number and position variance  are not ``complementary'' (in the QM sense)
since there is, of course, no uncertainty principle involved for classical
variables. Finally we observe that after convolution the laser rays do not
travel in a unique direction. However we shall refer to the laser direction as
the direction of the ray of peak intensity. This is the $z$-axis in
Eq.\,$(\ref{l2})$ .

\section*{\normalsize III. GEOMETRY OF A PROPOSED EXPERIMENT}

We sketch below the geometry of a laser source at $O$, with the
$z$-axis extended to  a dielectric block with normal the
$z_*$-axis. The angle between the $z$ and $z_*$ axis is $\alpha$.
A laser impinging along the $z$-axis travels, initially, within
the dielectric with refractive index greater than air in a
different direction, $z_d$, with angle $\beta(< \alpha)$ w.r.t the
$z_*$-axis and angle $\alpha-\beta$ w.r.t. the $z$-axis. The
outgoing situation when the laser exits the dielectric is, as we
shall see, conditioned by the impact angle and the laser spatial
dimensions. The axis indicated by $y_*$ and $z_*$, and for
completeness $x_*$, are obtained from $y$ and $z$, and $x$, by a
rotation of $\alpha$ around the $x$-axis. We chose the $z$-axis as
that of the center of the laser beam so that we may use directly
the expressions for the laser of the previous section. Since the
dielectric block is perpendicular to the $z_*$-axis, it is
stratified in the $z_*$ direction.
\begin{center}
    \setlength{\unitlength}{1mm}
    \begin{picture}(120,70)
        \put(0,5){\line(1,0){120}}
        \put(0,5){\line(0,1){60}}
        \put(120,5){\line(0,1){60}}
        \put(0,65){\line(1,0){120}}
\thicklines \put(10,35){\vector(1,-2){5}}
 \put(10,35){\vector(2,1){10}}
 \put(10,35){\vector(0,1){12}}
  \put(9.5,34){$\bullet$}
 \put(23,34){\large $\boldsymbol{z}$}
  \put(9,50){\large $\boldsymbol{y}$}
  \put(22,40){\large $\boldsymbol{y_*}$}
  \put(15,22){\large $\boldsymbol{z_*}$}

\thicklines   \put(10,35){\vector(1,0){12}}
 \put(14,17){\line(2,1){80}}
\put(19,7){\line(2,1){80}} \put(14,17){\line(1,-2){5}}
\put(94,57){\line(1,-2){5}}
\put(13,31){\large $\boldsymbol{\alpha}$} \put(54,33.5){\large
$\boldsymbol{\frac{\pi}{2}-\beta}$}  \put(45,36){\large $\boldsymbol{\alpha}$}


\thinlines \put(11,25){\vector(1,-2){3.2}} \put(11,25){\vector(-1,2){4.2}}
 \put(6.5,25){\large $\boldsymbol{a}$}
\put(98,53){\vector(1,-2){2.7}} \put(98,53){\vector(-1,2){2.4}}
 \put(99.5,53){\large $\boldsymbol{b}$}

 \multiput(32,34)(1,0){18}{$\cdot$}
 \multiput(50,34)(1,-1){15}{$\cdot$}
\put(48,40){\line(1,-2){5}}

\put(79.1,42.9){\Large $\boldsymbol{n}$}
 \put(3,58){\large \bf Air}
 \put(30,50){\large $\boldsymbol{\sin \alpha = n\,\sin \beta}$}
  \put(60,8){\large {\bf   Dielectric Layer Geometry}}
   \end{picture}
\end{center}

If a plane wave traverses the block, the separation of variables implies that
the wave numbers in the $x_*(=x)$ and $y_*$ directions remain unaltered. Only
the $z_*$ wave number changes, in analogy to the momentum in one-dimensional
QM, thus within the dielectric we have the following plane wave
\[
\exp[\,i\,( k_{x_*}x_*+k_{y_*}y_*+ q_{z_*}z_*\,)\,]\,\,.
\]
Such a plane wave satisfies the Maxwell equations for a stratified media,
\begin{equation}
\left(\,\partial^{^{2}}_{x_*}+ \partial^{^{2}}_{y_*} +
\partial^{^{2}}_{z_*} \right)\,E + n^{\2}(z_*\,)\,k^{\2}E
=0\,\,.
\end{equation}
From this equation we get
\[
q_{z_*} = \sqrt{n^{\2}k^{^{2}} - k^{^{2}}_{x_*}- k^{^{2}}_{y_*}}=
\sqrt{(n^{\2}-1)k^{^{2}} + k^{^{2}}_{z_*}} \,\,.
\]
Note that for $n>1$, $q_{z_*} $ is a {\em real} number. Thus, in our setup,
for whatever incident angle there are no evanescent waves and hence no
tunneling. We have only diffusion in QM terms. This implies that our
dielectric block corresponds to a potential well. Now we observe that, after
convolution with $G(k_x,k_y)$,
\[
k_{z_*}=k_z\,\cos\alpha - k_y\,\sin\alpha \approx k\,\cos\alpha -
k_y\,\sin\alpha\,\,,
\]
where the approximation is justified if we neglect terms of second order in
$k_x$ and $k_y$ and consequently spreading effects. Similarly we find, after
using Snell's law $\sin \alpha=n\,\sin\beta$,
\begin{equation}
q_{z_*} \approx n\, k \cos \beta - k_y\,\tan\beta \cos\alpha\,\,.
\end{equation}
Using these approximations and the fact that
$\boldsymbol{k}\cdot\boldsymbol{r} = \boldsymbol{k_*}\cdot\boldsymbol{r_*}$,
the plane wave phase within the dielectric,
\[k_{x_*}\,x_*+k_{y_*}\,y_*+q_{z_*}\,z_*\,\,,
\]
 can be rewritten as follows
\[
\begin{array}{l} k_x\,x+k_y\,y+k_z\,z
+(q_{z_*}-k_{z_*})\,z_*\\
 \approx\,\,
 k_x\,x+k_y\,\left[\,y+(\sin\alpha-\tan\beta\cos\alpha)\,z_*\,\right]
 + k\,\left[\,z+(n\,\cos\beta-\cos \alpha)\,z_*\,\right] \\
  =\,\,k_x\,x+k_y\,\frac{\cos\alpha}{\cos\beta}\,
 \underbrace{[\,\cos(\alpha-\beta)\,y + \sin(\alpha-\beta)\,z\,]}_{y_d}
 + n\,k\, \underbrace{[\,\cos(\alpha-\beta)\,z - \sin(\alpha-\beta)\,y\,]}_{z_d}\,\,.
\end{array}
\]
However we should not simply convolute the above  plane wave with
$G(k_x,k_y)$. We have first to calculate the ```step'' reflection and
transmission amplitudes. Indeed there is a reflected beam at the interface.
Nevertheless, if the transmission amplitude at the interface is close to one
we can ignore reflection  and write
\begin{equation}
\label{ld}
 E_d(x,y,z)
  \approx  E_{\0}\,\exp[\,i\,n\,k\,z_d\,]\,\,
  \exp\left[\,-\, \frac{x^{^{2}}+(\cos\alpha/\cos\beta)^{^{2}} y_d^{^{2}}}{\mbox{w}_{\0}^{\2}}
\right]\,\,.
\end{equation}
As expected this is seen to represent a laser with principal direction along
the $z_d$-axis and with the transverse $y_d$ spread increased by
$\cos\beta/\cos\alpha$. This deforms the laser cross section from circular to
elliptic. This result can also be directly derived by simple geometry and
Snell's law. Our analytic derivation confirms the validity of our
approximations which will also  be used in later sections.

 In the next section we calculate the reflection and
transmission coefficients  with the  method in which entry and exit from the
dielectric are treated separately as successive and repeated ``step''
calculations\cite{BDR04}.

\section*{\normalsize IV. TRANSMISSION AND REFLECTION AMPLITUDES}

As we have demonstrated in previous works\cite{DR05,DR06,DR08}, the
transmission and reflection amplitudes both in Optics and QM can be built up
by successive applications of the ``step'' analysis at dielectric or potential
discontinuities. In this case at dielectric interfaces. The general $z_*$
solutions, be they oscillatory or evanescent, are constrained by applying
continuity to the amplitude and its first derivative in $z_*$.

This procedure is repeated {\em ad infinitum} for multiply reflected
amplitudes. The final ``formal'' transmission and reflected amplitudes are
obtained by summing the appropriate series. However if, for example,
incoherence dominates between successive contributions then the individual
terms are more relevant than the formal sum. In numerical calculations, which
involve the convolution function and the plane wave as well, we must work, in
general, with the summed amplitudes to reproduce any, even partial, coherence
phenomena. Our two-step calculations are one-dimensional, $z_*$, and perfectly
recall those of QM.

\begin{center}
    \setlength{\unitlength}{1mm}
    \begin{picture}(120,70)
        \put(0,5){\line(1,0){120}}
        \put(0,5){\line(0,1){60}}
        \put(120,5){\line(0,1){60}}
        \put(0,65){\line(1,0){120}}
 \put(4.5,34){$\bullet$} \put(35,34){$\bullet$}
\thicklines   \put(5,35){\line(1,0){30}}

\multiput(5,34)(1,0){31}{$\cdot$} \multiput(5,33)(1,0){31}{$\cdot$}
\multiput(8,32)(1,0){28}{$\cdot$} \multiput(8,31)(1,0){28}{$\cdot$}
\multiput(5,30)(1,0){31}{$\cdot$} \multiput(5,29)(1,0){31}{$\cdot$}
\multiput(5,28)(1,0){31}{$\cdot$} \multiput(5,27)(1,0){31}{$\cdot$}
\multiput(5,26)(1,0){31}{$\cdot$} \multiput(5,25)(1,0){31}{$\cdot$}
\multiput(5,24)(1,0){31}{$\cdot$} \multiput(5,23)(1,0){31}{$\cdot$}
\multiput(5,22)(1,0){31}{$\cdot$} \multiput(5,21)(1,0){31}{$\cdot$}
\multiput(5,20)(1,0){31}{$\cdot$} \multiput(5,19)(1,0){31}{$\cdot$}
\multiput(5,18)(1,0){31}{$\cdot$} \multiput(5,17)(1,0){31}{$\cdot$}
\multiput(5,16)(1,0){31}{$\cdot$} \multiput(5,15)(1,0){31}{$\cdot$}

\put(44.5,34){$\bullet$} \put(75,34){$\bullet$}

\thicklines \put(45,35){\line(1,0){30}}

\multiput(45,35)(1,0){31}{$\cdot$} \multiput(48,36)(1,0){28}{$\cdot$}
\multiput(48,37)(1,0){28}{$\cdot$} \multiput(45,38)(1,0){31}{$\cdot$}
\multiput(45,39)(1,0){31}{$\cdot$} \multiput(45,40)(1,0){31}{$\cdot$}
\multiput(45,41)(1,0){31}{$\cdot$} \multiput(45,42)(1,0){31}{$\cdot$}
\multiput(45,43)(1,0){31}{$\cdot$} \multiput(45,44)(1,0){31}{$\cdot$}
\multiput(45,45)(1,0){31}{$\cdot$} \multiput(45,46)(1,0){31}{$\cdot$}
\multiput(45,47)(1,0){31}{$\cdot$} \multiput(45,48)(1,0){31}{$\cdot$}
\multiput(45,49)(1,0){31}{$\cdot$} \multiput(45,50)(1,0){31}{$\cdot$}
\multiput(45,51)(1,0){31}{$\cdot$} \multiput(45,52)(1,0){31}{$\cdot$}
\multiput(45,53)(1,0){31}{$\cdot$} \multiput(45,54)(1,0){31}{$\cdot$}
\multiput(45,55)(1,0){31}{$\cdot$}

\put(84.5,34){$\bullet$} \put(115,34){$\bullet$} \thicklines
\put(85,35){\line(1,0){30}}

\multiput(85,34)(1,0){31}{$\cdot$} \multiput(85,33)(1,0){31}{$\cdot$}
\multiput(88,32)(1,0){28}{$\cdot$} \multiput(88,31)(1,0){28}{$\cdot$}
\multiput(85,30)(1,0){31}{$\cdot$} \multiput(85,29)(1,0){31}{$\cdot$}
\multiput(85,28)(1,0){31}{$\cdot$} \multiput(85,27)(1,0){31}{$\cdot$}
\multiput(85,26)(1,0){31}{$\cdot$} \multiput(85,25)(1,0){31}{$\cdot$}
\multiput(85,24)(1,0){31}{$\cdot$} \multiput(85,23)(1,0){31}{$\cdot$}
\multiput(85,22)(1,0){31}{$\cdot$} \multiput(85,21)(1,0){31}{$\cdot$}
\multiput(85,20)(1,0){31}{$\cdot$} \multiput(85,19)(1,0){31}{$\cdot$}
\multiput(85,18)(1,0){31}{$\cdot$} \multiput(85,17)(1,0){31}{$\cdot$}
\multiput(85,16)(1,0){31}{$\cdot$} \multiput(85,15)(1,0){31}{$\cdot$}

\put(4.5,34){$\bullet$} \put(35,34){$\bullet$} \thicklines
\put(5,35){\line(1,0){30}}


\thicklines \put(5,50){\line(1,-1){15}} \put(20,35){\line(1,1){15}}
\put(60,35){\line(1,-1){15}} \put(100,35){\line(1,1){15}}
\put(5,50){\vector(1,-1){9}} \put(20,35){\vector(1,1){8}}
\put(60,35){\vector(1,-1){8}} \put(100,35){\vector(1,1){8}}

\put(20,35){\line(1,-3){6.5}} \put(53,56){\line(1,-3){7}}
\put(60,35){\line(1,3){7}} \put(100,35){\line(1,-3){6.5}}
\put(93.2,15.5){\line(1,3){6.5}}

\put(20,35){\vector(1,-3){3.5}} \put(53,56){\vector(1,-3){3.8}}
\put(60,35){\vector(1,3){4.2}} \put(100,35){\vector(1,-3){4}}
\put(93.2,15.5){\vector(1,3){3.3}}

\put(4,51){\large $\boldsymbol{1}$} \put(35,51){\large $\boldsymbol{R_{\1}}$}
 \put(25.5,11){\large $\boldsymbol{T_{\1}}$}

\put(52,57){\large $\boldsymbol{1}$} \put(66,57){\large $\boldsymbol{R_{\2}}$}
 \put(75,16){\large $\boldsymbol{T_{\2}}$}

 \put(92,10.5){\large $\boldsymbol{1}$} \put(105,10.5){\large $\boldsymbol{\widetilde{R}_{\1}}$}
 \put(115,51){\large $\boldsymbol{\widetilde{T}_{\1}}$}

\put(5,31.4){\large $\boldsymbol{n}$} \put(45,36.4){\large $\boldsymbol{n}$}
\put(85,31.4){\large $\boldsymbol{n}$}

\put(3,58){\large \bf Air}

\put(35,8){\large {\bf   Multiple Step Analysis}}

   \end{picture}
\end{center}
\noindent The above trilogy of figures are those relevant to a plane wave
impinging upon a dielectric block. In the first figure (step $1\to n$)
$R_{\1}$ is the reflected amplitude and $T_{\1}$ the transmitted amplitude
within the dielectric for an incident plane wave of unit amplitude. The
continuity equations when solved yield
\begin{equation}
R_{\1}=\displaystyle{\frac{k_{z_*} -\,q_{z_*} }{k_{z_*}+\,q_{z_*}}}\,
\exp\left[\,2\,i\,k_{z_*}\,a\,\right]\,\,\,,\,\,\,\,\,\,\,
T_{\1}=2\,\displaystyle{\frac{k_{z_*}}{k_{z_*}+\,q_{z_*}}}\, \exp\left[\,i\,(
k_{z_*}- q_{z_*})\,a\,\right]\,\,.
\end{equation}
Note that, in accordance with flux conservation\cite{Born99},
\[ |R_{\1}|^{^{2}}+\,\frac{q_{z_*}}{k_{z_*}}\,\,|T_{\1}|^{^{2}}=1\,\,.\]
 The second figure (step $n\to 1$) corresponds to the exiting of the
plane wave from the dielectric. Again it is convenient to use a unit
normalization for the incoming wave. The results for the reflected, $R_{\2}$,
and transmitted, $T_{\2}$, amplitudes are given by the previous expressions
with $k_{z_*}\leftrightarrow \,\,q_{z_*}$ and $a\to a+b$,
\begin{equation}
R_{\2}=\displaystyle{\frac{q_{z_*} -\,k_{z_*} }{k_{z_*}+\,q_{z_*}}}\,
\exp\left[\,2\,i\,q_{z_*}\,(a+b)\,\right]\,\,\,,\,\,\,\,\,\,\,
T_{\2}=2\,\displaystyle{\frac{q_{z_*}}{k_{z_*}+\,q_{z_*}}}\, \exp\left[\,i\,(
q_{z_*}- k_{z_*})\,(a+b)\,\right]\,\,.
\end{equation}
Again we observe that, as required,
\[ |R_{\2}|^{^{2}}+\,\frac{k_{z_*}}{q_{z_*}}\,\,|T_{\2}|^{^{2}}=1\,\,.\]
Finally a reflected wave at the second interface, $z_*=a+b$, returns to the
first interface, $z_*=a$, and again gives rise to a reflected,
$\widetilde{R}_{\1}$, and transmitted, $\widetilde{T}_{\1}$, wave. The
solution of the continuity equations are now obtained from $R_{\1}$ and
$T_{\1}$ by $k_{z_*}\leftrightarrow \,\,-\,q_{z_*}$,
\begin{equation}
\widetilde{R}_{\1}=\displaystyle{\frac{q_{z_*} -\,k_{z_*}
}{k_{z_*}+\,q_{z_*}}}\,
\exp\left[\,-\,2\,i\,q_{z_*}\,a\,\right]\,\,\,,\,\,\,\,\,\,\,
\widetilde{T}_{\1}=2\,\displaystyle{\frac{q_{z_*}}{k_{z_*}+\,q_{z_*}}}\,
\exp\left[\,i\,( k_{z_*}- q_{z_*})\,a\,\right]\,\,.
\end{equation}
These amplitudes satisfy
\[
|\widetilde{R}_{\1}|^{^{2}}+\,\frac{k_{z_*}}{q_{z_*}}\,\,
|\widetilde{T}_{\1}|^{^{2}}=1\,\,.
\]
The total transmission and reflection amplitudes can now be built up by
repeated multiplication and summation of the reflection and transmission
coefficients just given. The figure below gives the first few of these
multiple reflection and transmission terms.
\begin{center}
    \setlength{\unitlength}{1mm}
    \begin{picture}(120,70)
        \put(0,5){\line(1,0){120}}
        \put(0,5){\line(0,1){60}}
        \put(120,5){\line(0,1){60}}
        \put(0,65){\line(1,0){120}}
  \put(5.5,34){$\bullet$}

\thicklines   \put(6,35){\line(1,0){12}}
 \put(14,17){\line(2,1){80}}
\put(19,7){\line(2,1){80}} \put(14,17){\line(1,-2){5}}
\put(94,57){\line(1,-2){5}}

\multiput(14.5,16)(2,1){40}{$\cdot$} \multiput(15,15)(2,1){40}{$\cdot$}
\multiput(15.5,14)(2,1){40}{$\cdot$} \multiput(16,13)(2,1){40}{$\cdot$}
\multiput(16.5,12)(2,1){40}{$\cdot$} \multiput(17,11)(2,1){40}{$\cdot$}
\multiput(17.5,10)(2,1){40}{$\cdot$} \multiput(18,9)(2,1){40}{$\cdot$}
\multiput(18.5,8)(2,1){40}{$\cdot$} \multiput(19,7)(2,1){40}{$\cdot$}

\thinlines \put(11,25){\vector(1,-2){3.5}} \put(11,25){\vector(-1,2){4.5}}
 \put(6.5,25){\large $\boldsymbol{a}$}
\put(98,53){\vector(1,-2){2.7}} \put(98,53){\vector(-1,2){2.4}}
 \put(99.5,53){\large $\boldsymbol{b}$}

\thicklines \put(6,35){\line(1,0){44}}

\put(68,44){\line(1,1){13}} \put(62,41){\line(1,1){16}}
\put(56,38){\line(1,1){19}} \put(50,35){\line(1,1){22}}

\put(68,44){\vector(1,1){6}} \put(62,41){\vector(1,1){9}}
\put(56,38){\vector(1,1){12}} \put(50,35){\vector(1,1){15}}

\put(68,44){\line(1,-5){2.3}}\put(68,44){\line(1,-1){5}}
\put(62,41){\line(1,-5){2.3}}\put(62,41){\line(1,-1){8.4}}
\put(56,38){\line(1,-5){2.3}}\put(56,38){\line(1,-1){8.4}}
\put(50,35){\line(1,-1){8.4}}

\put(70.5,32.6){\vector(1,0){18}} \put(64.5,29.6){\vector(1,0){24}}
\put(58.5,26.6){\vector(1,0){30}}

 \put(70.5,32.6){\line(1,0){33}}
\put(64.5,29.6){\line(1,0){39}} \put(58.5,26.6){\line(1,0){45}}

 \put(6,35){\vector(1,0){22}}

 \put(3,60){\bf {\large Reflection:}  $\boldsymbol{R_{\1}}$, $
 \boldsymbol{T_{\1}R_{\2}\widetilde{T}_{\1}}$,
$\boldsymbol{T_{\1}R_{\2}\widetilde{R}_{\1}R_{\2}\widetilde{T}_{\1}}$,
$\boldsymbol{T_{\1}R_{\2}(\widetilde{R}_{\1}R_{\2})^{^{2}}\widetilde{T}_{\1}}$,
...}

\put(32,8){\bf {\large Transmission:}  $\boldsymbol{T_{\1}T_{\2}}$,
$\boldsymbol{T_{\1}R_{\2}\widetilde{R}_{\1}T_{\2}}$,
$\boldsymbol{T_{\1}(R_{\2}\widetilde{R}_{\1})^{^{2}}T_{\2}}$, ...}
   \end{picture}
\end{center}
The first contribution to the total reflection amplitude is simply $R_{\1}$.
The second contribution is clearly $T_{\1}R_{\2}\widetilde{T}_{\1}$. The third
is $T_{\1}R_{\2}\widetilde{R}_{\1}R_{\2}\widetilde{T}_{\1}$ and so forth.
Hence the full reflection amplitude is
\begin{equation}
R= R_{\1} +
T_{\1}R_{\2}\sum_{m=0}^{\infty}(\widetilde{R}_{\1}R_{\2})^{^{m}}\widetilde{T}_{\1}=
R_{\1} + \frac{T_{\1}R_{\2}\widetilde{T}_{\1}}{1-\widetilde{R}_{\1}R_{\2}} =
\frac{i\,(q_{z_{*}}^{^{2}}-\,k_{z_{*}}^{^{2}})\,\sin(q_{z_{*}}b)\,e^{2\,i\,k_{z_{*}}a}}{
2\,k_{z_{*}}q_{z_{*}} \cos(q_{z_{*}}b) -
i\,(k_{z_{*}}^{^{2}}+\,q_{z_{*}}^{^{2}})\sin(q_{z_{*}}b)} \,\,.
\end{equation}
In the same way, it is easy to derive the full transmitted amplitude
\begin{equation}
T= T_{\1}\sum_{m=0}^{\infty}(\widetilde{R}_{\1}R_{\2})^{^{m}}T_{\2}=
\frac{T_{\1}T_{\2}}{1-\widetilde{R}_{\1}R_{\2}}=\frac{2\,k_{z_{*}}q_{z_{*}}\,e^{-\,i\,k_{z_{*}}b}}{
2\,k_{z_{*}}q_{z_{*}} \cos(q_{z_{*}}b) -
i\,(k_{z_{*}}^{^{2}}+\,q_{z_{*}}^{^{2}})\sin(q_{z_{*}}b)}\,\,.
\end{equation}
Now these amplitudes, after some calculation, satisfy the well known
condition\cite{Born99,Cohen77}
\[|R|^{^{2}}+|T|^{^{2}}=1\,\,,\]
as must be the case when total coherence and a single reflected and
transmitted wave in free space occur. However we also find that the sum of the
modulus squared of all the contributions to reflection,
\[ |R_{\1}|^{^{2}} +
\frac{|T_{\1}R_{\2}\widetilde{T}_{\1}|^{^{2}}}{1-|\widetilde{R}_{\1}R_{\2}|^{^{^{2}}}}=
\frac{(k_{z_*}-\,q_{z_*})^{^{2}}}{k^{^{2}}_{z_*}+\,q^{^{2}}_{z_*}} \,\,,
\]
and to transmission,
\[\frac{|T_{\1}T_{\2}|^{^{2}}}{1-|\widetilde{R}_{\1}R_{\2}|^{^{^{2}}}}=
\frac{2\,k_{z_*}q_{z_*}}{k^{^{2}}_{z_*}+\,q^{^{2}}_{z_*}}\,\,,
\]
also sum to one. This result is reasonable since it implies conservation of
intensity in the limit of total incoherence. Obviously a similar result must
hold for any level of coherence. We recall that the transmission and
reflection intensities are crucially dependent upon the degree of coherence.
For example, total coherence displays the phenomena of {\em
resonance}\cite{Cohen77}.

\section*{\normalsize V. TRANSMISSION PHENOMENA}

In this section we calculate both numerically and with the use of the SPM the
exit positions of individual beams for various incident angles of our gaussian
laser. For these calculations, we have chosen
\[\begin{array}{lclclcl}
\lambda & = & 633\,\mbox{nm}\,\,\,,& \,\,\,\,\,\,\, &
\mbox{w}_{\0}=2\,\mbox{mm}\,\,\,,\\
a & = & 10\,\mbox{cm}\,\,\,,& \,\,\,\,\,\,\, &
b=5\,\mbox{cm}\,\,\,,\\
n & = & \sqrt{3}\,\,\,,& \,\,\,\,\,\,\, &
\alpha=\pi/60\,\,,\pi/30\,\,,\pi/6\,\,,\pi/3\,\,\,.
\end{array}
\]
The expression for the transmitted amplitude is, with the approximation
described in the text for $k_z$,
\begin{equation}
\label{lasT}
 E_{\T}(x,y,z) = E_{\0}\,
\frac{w_{\0}^{^{2}}}{4\,\pi}\,\,\int \,
\mbox{d}k_x\,\mbox{d}k_{y}\,\,T(k_x,k_y)\,G(k_x,k_y)\,\exp\left[\,i\,\left(k_x\,x+k_{y}\,y+k_z\,z\,\right)\,
\right]\,\,.
\end{equation}
Now, as shown in Table 1, the dependence of $T$ upon $k_x$ is almost
negligible. This follows from the fact that all angles are rotations about the
$x$-axis and thus do not involve $k_x$. On the other hand the angles do depend
upon $k_y$ to first order. Thus we can approximate $T(k_x,k_y)$ with
$T(0,k_y)$. Furthermore, the use of a sharp gaussian allows us to use
$k_z\approx k - (k_x^{^{2}}+k_y^{^{2}})/2k$. Consequently the $k_x$
integration can be performed analytically, as for the incoming laser, and we
obtain
\begin{eqnarray}
\label{lasTa} E_{\T}(x,y,z) &\approx&
E_{\0}\,\frac{\mbox{w}^{\2}_{\0}}{4\,\pi}\,\,e^{\,i\,k\,z}\,\,
\sqrt{\frac{4\,\pi}{\mbox{w}_{\0}^{\2}+2\,i\,z/\,k}}\, \exp\left[\,-\,
\frac{x^{^{2}}}{\mbox{w}_{\0}^{\2}+2\,i\,z/\,k} \right]\,\,\times \nonumber \\
 & &
\int \, \mbox{d}k_{y}\,\,T(0,k_y)\,\exp\left[\,-\,\left(\,
\frac{w_{\0}^{\2}}{4} + i \, \frac{z}{2\,k}\,\right)\, k_{y}^{^{2}}\,+\,i\,
k_{y}\,y \right]\,\,.
\end{eqnarray}
This integral must be performed numerically. In Fig.\,2, we plot the results
for $|E_{\T}/E_{\0}|$ versus $y$ for four values of the incoming $\alpha$. The
$x$-dependence is not plotted because it is the same as the incoming laser
beam.  We observe that for small $\alpha$ angles  a single beam emerges.
However, its peak is somewhat displaced w.r.t. the incoming beam (an effect of
$T$). As $\alpha$ increases, evidence of a secondary beam is seen. For even
high $\alpha$, this beam and others separate and we observe the effect of the
incoherent limit of the transmission coefficient. Our curves are calculated
directly from the full $T$, however they of course agree with those calculated
from each appropriate transmission term. We exhibit only the first three terms
in our plots.

In Fig.\,3 we plot these same results in a different manner suitable to direct
experimental verification. We plot the $xy$-profile of each intensity for the
four chosen $\alpha$. For these plots we have arbitrarily chosen
$z=1\,\mbox{m}$, however, with the very small spreading in our simulation, the
exact value of $z$ (below several meters) is irrelevant. All centers lie upon
the $x$-axis as does the incoming beam profile. However the $y$ values depend
upon $\alpha$. Another way of determining these $y$ values without going
through a numerical calculation is by means of the SPM. We must now work with
the individual transmission terms since the SPM only yields a single peak
position $y$. The SPM  applied to an expression which involves a sum of peaks
yields only  the mean value, normally not even coincident with any of the
peaks. The SPM involves taking the derivative of the argument of the phase of
a peaked function assuming it more relevant to the peak's position compared to
the modulus of the amplitude. Now the phase of our amplitudes take the form
\[
e^{i\,[k_y y+ f(k_y)]}\,\,,
\]
where $f(k_y)$ is a function of $k_y$ (independent of any $y$ dependence). The
function $f$ is calculated below. The SPM then says that the peak occurs where
\[ y + \left.\frac{\mbox{d}f(k_y)}{\mbox{d}k_y}\right|_{_{k_y=0}}=\,\,0\,\,.\]
Consider, the first leading transmission amplitude term $T_{\1}T_{\2}$, the
phase is
\[(q_{z_*}-\,k_{z_*})(a+b)\,+\,(k_{z_*}-\,q_{z_*})\,a = (q_{z_*}-\,k_{z_*})\,b\,\,, \]
to which, we must add the plane wave phase $e^{ik_yy}$. Thus in this case,
$f(k_y)=(q_{z_*}-\,k_{z_*})\,b$. Before deriving w.r.t. $k_y$ this function,
we recall that
\[q_{z_*}-\,k_{z_*} = k\,(n\cos \beta - \cos \alpha)+ k_y\,(\sin \alpha -
\tan\beta \cos\alpha) + \mbox{O}\left(k_y^{^{2}}\right)\,\,.\]
Thus,
\[
\left.\frac{\mbox{d}f(k_y)}{\mbox{d}k_y}\right|_{_{k_y=0}}
= \,\,\frac{\sin(\alpha-\beta)}{\cos\beta}\,\,b\,\,.
\]
Hence, for the first transmission term, the exit point is
\[y_{_{(1)}}[\,z_*=a+b\,]= -\,\frac{\sin(\alpha-\beta)}{\cos\beta}\,\,b\,\,.\]
The linear dependence upon $b$ is obvious because of geometry, indeed this
expression and those below are also derivable by pure geometry. Repeating this
procedure for the $m$-th peak ($m=1$ being the first), we must include the
phase of $(R_{\2}\widetilde{R}_{\1})^{^{m-1}}$. This yields
\begin{equation}
y_{_{(m)}}[z_*=a+b]= -\, \left[\, \frac{\sin(\alpha-\beta)}{\cos\beta}
-2\,(m-1)\,
 \tan \beta \cos\alpha\,\right]\,b\,\,.
 \end{equation}
In particular, the resulting values of the peaks for $\alpha=\pi/3$ are
$y_{_{(m)}}= -\,(2-m)\,b\,/\,\sqrt{3}$.  These agree very well with the center
positions plotted in Fig.\,3. The integrated power $P$ of each outgoing laser
beam {\em relative} to the incoming beam power is given by
\begin{equation}
\label{prel}
 P_{rel} \approx \frac{\mbox{w}_{\0}}{2\sqrt{2}\,\pi^{^{3/2}}}\,\int \,
 \mbox{d}y\,\,\left|\,\int\,
 \mbox{d}k_{y}\,\,T(0,k_y)\,\exp\left[\,-\,\left(\,
\frac{w_{\0}^{\2}}{4} + i \, \frac{z}{2\,k}\,\right)\, k_{y}^{^{2}}\,+\,i\,
k_{y}\,y \right]\,\right|^{^{2}}\,\,.
\end{equation}
For the calculation of the separate laser terms it is sufficient to use the
following approxiamtions
\[
|\,T_{\1}T_{\2}\,|  \approx  4\,n\cos\alpha\,\cos\beta\,\mbox{\large
$/$}(\cos\alpha+n\cos\beta)^{^{2}}\,\,\,\,\,\,\,\mbox{and}\,\,\,\,\,\,\,
|\,R_{\2}\widetilde{R}_{\1}\,| \approx [\,(\cos\alpha-n\cos\beta)\mbox{\large
$/$}(\cos\alpha+n\cos\beta)\,]^{^{2}}\,\,.
\]
These are valid for small $k_{x,y}$. For example, for the particular case of
an incident angle $\alpha=\pi/3$ and for a refractive index $n=\sqrt{3}$ (for
which we have a diffusion angle $\beta=\pi/6$),
\[
|\,T_{\1}T_{\2}\,|\approx3/4\,\,\,\,\,\,\,\mbox{and}\,\,\,\,\,\,\,|\,R_{\2}\widetilde{R}_{\1}\,|
\approx 1/4\,\,.
\]
Hence the first three laser transmission beams have the following relative
powers
\[
P_{rel}(\pi/3)\approx\left(\,\frac{9}{16}\,,\,\frac{9}{16^2}\,,\,
\frac{9}{16^3}\,\right)\,\,.
\]
This result is in excellent agreement with the numerical results obtained from
Eq.(\ref{prel}) and shown in Fig.\,3.

\section*{\normalsize VI. CONCLUSIONS}

 Our principal objective in this paper is to develop the formalism,
 conventions and approximations for the interaction of a gaussian laser
 beam with dielectric structures. We emphasize in particular the dominant role
 played by the momentum perpendicular to the dielectric block, $z_*$
 component, and in general to any stratified dielectric system. This result
 draws a strong analogy with one-dimensional non-relativistic QM. Indeed we
 have noted in the text that our system corresponds to a QM well potential.
 The numerical and other calculations displayed in this work invite
 experimental confirmation. It is particularly interesting to note the
 conditions for coherence and incoherence in this study and compare them with
 those of one-dimensional QM. In QM we deal with time dependent wave packets
 and coherence requires the size of these wave packets to be large w.r.t. the
 well size. In this work the analogy is the separation in the $y$-variable
 of laser beams. For this optical study, coherence can be achieved not only by
 increasing the laser beam size w.r.t. the block depth $b$ so that overlap
 occurs between outgoing beams, but also by reducing the incident angle
 $\alpha$. This also leads to overlap of individual terms since the separation
 of the $y$ exit points is reduced, for any fixed laser size, by decreasing
 $\alpha$. We have calculated the various $y$ values for incoherence, both by
 the SPM and numerically and found complete agreement. They remain, of course,
 to be confirmed experimentally.
We have also calculated, for various incident angles $\alpha$, the outgoing
incoherent beam intensities and profiles.

A more complex and interesting situation occurs if we consider a dielectric
slab cut diagonally, see the figure below.

\begin{center}
    \setlength{\unitlength}{1mm}
    \begin{picture}(45,80)
        \thicklines
        \put(2,70){\line(1,0){40}}
        \put(42,70){\line(0,-1){40}}
        \put(2,70){\line(1,-1){40}}

        \put(2,60){\line(0,-1){40}}
        \put(42,20){\line(-1,0){40}}
        \put(2,60){\line(1,-1){40}}

       \put(0,75){\line(2,-1){10}} \put(0,75){\vector(2,-1){5}}
       \put(10,70){\line(1,-2){8}}  \put(10,70){\vector(1,-2){4}}
       \put(10,70){\line(2,1){20}}   \put(10,70){\vector(2,1){10}}
       \put(18,54){\line(2,-1){14}} \put(18,54){\vector(2,-1){8}}

        \thinlines
         \multiput(18.5,54)(-1,-1){6}{$\cdot$}
          \put(37,66){{\large $\boldsymbol{n}$}}
           \put(3.5,21.5){{\large $\boldsymbol{n}$}}

           \put(35,74){\large \bf Air}

 \thicklines

           \put(17.25,53){$\bullet$}
           \put(19.25,41){$\bullet$}

         \put(20,42){\line(1,-2){11}} \put(20,42){\vector(1,-2){6}}
         \put(31,20){\line(2,-1){12}} \put(31,20){\vector(2,-1){7}}
         \put(31,20){\line(1,2){2.5}} \put(31,20){\vector(1,2){1.75}}

         \multiput(9.5,70)(0,1){7}{$\cdot$}
         \put(6.3,72.5){$\boldsymbol{\alpha}$}
         \multiput(9.5,70)(0,-1){9}{$\cdot$}
         \put(13.1,66.2){$\boldsymbol{\frac{\pi}{2}-\beta}$}
         \put(38.95,34){$\boldsymbol{\frac{\pi}{4}}$}
         \multiput(18.5,54)(1,1){7}{$\cdot$}
         \put(17.3,56.7){$\boldsymbol{\varphi}$}

\end{picture}
\end{center}

\vspace*{-1.2cm}

\noindent The dielectric-air-dielectric system now mimics a
potential barrier and in addition to diffusion we encounter
tunneling. Tunneling still holds a number of theoretical
conundrums notwithstanding its practical importance in technology
such as in the tunneling microscope. It seems that technology has
leap-frogged theory in this case. For example, there is still
discussion in the literature about tunneling
times\cite{HAR,Ref7,Ref8} and possible superluminal
velocities\cite{ORJ04,W06}. Another theoretical conundrum is that
for tunneling, our series calculation (two step method) yields a
non convergent sum. Transmission seems always to be coherent,
independent of the barrier/wave packet ratio. In the optical
analogy, we pass from diffusion to tunneling by simply varying the
incident angle. Tunneling occurs when $\varphi>\beta_{\mbox{\tiny
critical}}=\arcsin(1/n)$.

Let us rapidly recall some of the contents of the previous sections. We
started by representing the analytic function of a gaussian laser\cite{EP04}
and consequently extended our studies to 3-dimensional optics. We also
extended this analysis to the laser beam within the dielectric. We then
presented the optical analogy of a QM well problem and derived both by
numerical and approximate analytic methods, SPM\cite{BH75}, the exit points
from the dielectric block of the laser beam or beams for various incidence
angles. We have calculated the intensities of the first few incoherent terms
(beams) in the transition amplitude for a realistic experimental setup.

Finally, in this section, we have outlined a structure suitable
for investigating both diffusion and tunneling again dependent
upon the incident angle. Another more complex structure would be
the optical analogy of twin barrier tunneling\cite{DR05} which, as
known, exhibits the extraordinary phenomena of resonant
tunneling\cite{Ref3,ResT}. These structures will be the subject of
a subsequent report.

\section*{\small \rm ACKNOWLEDGMENTS}

One of the authors (S.D.L.) gratefully acknowledges CNPq/Fapesp
(Brazil) for financial support. The author also thanks the
University of Salento (Lecce, Italy)  for the invitation and the
hospitality.

\newpage

\begin{table}
\begin{center}
\begin{tabular}{|r|c|c|c|c|}
\hline
\multicolumn{5}{|c|}{$\boldsymbol{T(k_x,k_y)}$} \\
\hline \hline $\boldsymbol{\alpha=\pi/60}\,\,\bullet$
 & $\mbox{w}_{\0}k_x\,:\,\,\,\, 0\hspace*{1.2cm}$ & $\pm 2$ & $\pm 4$ & $\pm 6 $
\\  \hline \hline
$\mbox{w}_{\0}k_y\,:\,\,\, -6$ & (\,0.380\,,\,0.859\,) & (\,0.380\,,\,0.860\,) & (\,0.377\,,\,0.862\,) & (\,0.373\,,\,0.865\,) \\
\hline
$-4$ & (\,-0.475\,,\,0.788\,) & (\,-0.476\,,\,0.788\,) &(\,-0.478\,,\,0.786\,) & (\,-0.482\,,\,0.782\,) \\
\hline $-2$ & (\,-0.900\,,\,0.200\,) & (\,-0.900\,,\,0.201\,) &
(\,-0.900\,,\,0.204\,) & (\,-0.900\,,\,0.209\,) \\ \hline $\hspace*{.25cm}0$ &
(\,-0.356\,,\,0.868\,) & (\,-0.355\,,\,0.868\,) & (\,-0.352\,,\,0.869\,) &
(\,-0.347\,,\,0.869\,) \\ \hline
 $2$ & (\,0.686\,,\,-0.592\,) & (\,0.687\,,\,-0.592\,) &
 (\,0.689\,,\,-0.590\,)& (\,0.693\,,\,-0.586\,)
 \\ \hline
$4$ & (\,0.921\,,\,0.248\,) &  (\,0.920\,,\,0.249\,) &  (\,0.919\,,\,0.252\,)
& (\,0.917\,,\,0.256\,) \\ \hline $6$ & (\,0.060\,,\,0.892\,) &
(\,0.059\,,\,0.892\,) & (\,0.057\,,\,0.893\,) & (\,0.051\,,\,0.894\,)
 \\ \hline \hline
 $\boldsymbol{\alpha=\pi/30}\,\,\bullet$
 & $\mbox{w}_{\0}k_x\,:\,\,\,\, 0\hspace*{1.2cm}$ & $\pm 2$ & $\pm 4$ & $\pm 6 $
\\  \hline \hline
$\mbox{w}_{\0}k_y\,:\,\,\, -6$ &  (\,0.345\,,\,0.904\,)    &
(\,0.344\,,\,0.904\,)&(\,0.341\,,\,0.904\,)
 & (\,0.337\,,\,0.905\,)\\ \hline $-4$ &
(\,-0.941\,,\,-0.284\,) & (\,-0.941\,,\,-0.285\,) & (\,-0.940\,,\,-0.287\,)&
(\,-0.938\,,\,-0.291\,)
\\
\hline $-2$ & (\,0.813\,,\,-0.572\,) &(\,0.814\,,\,-0.571\,) &
(\,0.815\,,\,-0.570\,)& (\,0.817\,,\,-0.565\,)\\ \hline $0$ &
(\,-0.056\,,\,0.998\,) & (\,-0.056\,,\,0.998\,) & (\,-0.059\,,\,0.998\,) &
(\,-0.063\,,\,0.997\,) \\ \hline $2$ & (\,-0.746\,,\,-0.664\,)
 & (\,-0.746\,,\,-0.664\,)
& (\,-0.744\,,\,-0.666\,) & (\,-0.742\,,\,-0.670\,)
\\ \hline $4$ &
(\,0.978\,,\,-0.170\,)
&(\,0.978\,,\,-0.169\,) &(\,0.979\,,\,-0.167\,) & (\,0.980\,,\,-0.163\,)\\
\hline $6$ & (\,-0.469\,,\,0.862\,)& (\,-0.470\,,\,0.862\,)&
(\,-0.473\,,\,0.861\,)& (\,-0.477\,,\,0.859\,)\\ \hline \hline
$\boldsymbol{\alpha=\pi/6}\,\,\bullet$
 & $\mbox{w}_{\0}k_x\,:\,\,\,\, 0\hspace*{1.2cm}$ & $\pm 2$ & $\pm 4$ & $\pm 6 $
\\  \hline \hline
$\mbox{w}_{\0}k_y\,:\,\,\, -6$ &(\,0.161\,,\,0.942\,) &
(\,0.160\,,\,0.942\,)&(\,0.158\,,\,0.943\,) & (\,0.153\,,\,0.945\,)\\ \hline
$-4$ &(\,0.775\,,\,0.632\,) &(\,0.774\,,\,0.633\,)  & (\,0.773\,,\,0.634\,)
&(\,0.770\,,\,0.637\,)
\\ \hline $-2$ & (\,0.946\,,\,-0.024\,)
&  (\,0.946\,,\,-0.023\,)& (\,0.946\,,\,-0.020\,) &  (\,0.947\,,\,-0.016\,)\\ \hline $0$ &
 (\,0.689\,,\,-0.518\,)
 &  (\,0.689\,,\,-0.518\,) & (\,0.691\,,\,-0.516\,) &  (\,0.693\,,\,-0.511\,)\\ \hline $2$ &
 (\,0.301\,,\,-0.764\,)
  & (\,0.302\,,\,-0.764\,)& (\,0.305\,,\,-0.763\,)& (\,0.310\,,\,-0.761\,)\\ \hline $4$ &
  (\,-0.134\,,\,-0.832\,) &   (\,-0.134\,,\,-0.832\,)&   (\,-0.131\,,\,-0.833\,)&
   (\,-0.126\,,\,-0.836\,)\\
\hline $6$ &
  (\,-0.624\,,\,-0.674\,) &  (\,-0.623\,,\,-0.675\,) &  (\,-0.622\,,\,-0.678\,) & (\,-0.629\,,\,-0.681\,)  \\ \hline  \hline $\boldsymbol{\alpha=\pi/3}\,\,\bullet$
 & $\mbox{w}_{\0}k_x\,:\,\,\,\, 0\hspace*{1.2cm}$ & $\pm 2$ & $\pm 4$ & $\pm 6 $
\\  \hline \hline
$\mbox{w}_{\0}k_y\,:\,\,\, -6$ &
 (\,0.397\,,\,0.480\,)  & (\,0.397\,,\,0.480\,) &  (\,0.395\,,\,0.482\,)&  (\,0.393\,,\,0.484\,)\\
  \hline $-4$ &  (\,0.054\,,\,-0.997\,)&
  (\,0.055\,,\,-0.997\,) & (\,0.057\,,\,-0.997\,)& (\,0.060\,,\,-0.997\,)\\ \hline $-2$ &
  (\,-0.401\,,\,0.464\,)
&   (\,-0.401\,,\,0.464\,)&   (\,-0.402\,,\,0.462\,)&  (\,-0.404\,,\,0.461\,) \\ \hline $0$ &
  (\,0.670\,,\,-0.472\,) & (\,0.671\,,\,-0.472\,)  &  (\,0.671\,,\,-0.470\,) &
   (\,0.672\,,\,-0.466\,) \\ \hline $2$ &  (\,-0.665\,,\,-0.345\,) &  (\,-0.665\,,\,-0.346\,)&
    (\,-0.664\,,\,-0.348\,) &  (\,-0.664\,,\,-0.350\,)\\ \hline $4$ &
    (\,0.500\,,\,0.393\,)&   (\,0.500\,,\,0.393\,)&   (\,0.499\,,\,0.395\,)&  (\,0.496\,,\,0.397\,) \\
\hline $6$ &
  (\,-0.051\,,\,-0.983\,)
 &(\,-0.051\,,\,-0.983\,) & (\,-0.049\,,\,-0.983\,)& (\,-0.046\,,\,-0.984\,)\\ \hline
\end{tabular}
\end{center}
\caption{Values of the transmission amplitude, $(\mbox{Re}[T],\mbox{Im}[T])$,
as a function of $k_x$ and $k_y$ for four incident angles,
$\alpha=(\pi/60,\,\pi/30,\,\pi/6,\,\pi/3)$, for the chosen parameters
$\mbox{w}_{\0}=2\,\mbox{mm}$, $\lambda=633\,\mbox{nm}$, $n=\sqrt{3}$ and
$b=5\,\mbox{cm}$. The dependence of $T$ upon $k_x$ is clearly very small.}
\end{table}

\newpage

\begin{figure}[hbp]
\hspace*{-2.5cm}
\includegraphics[width=19cm, height=22cm, angle=0]{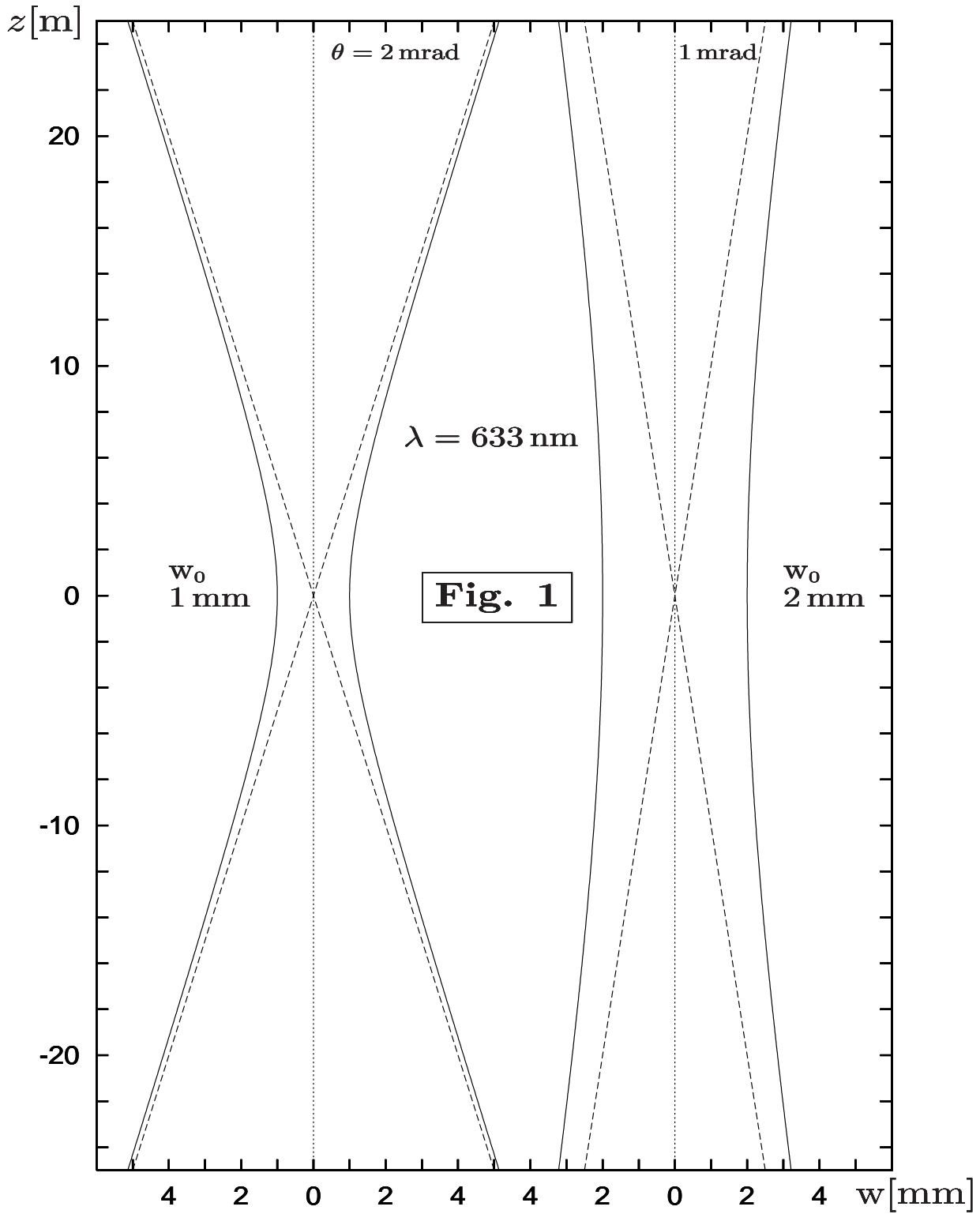}
\vspace*{-1.8cm}
 \caption{Laser intensity profiles in air for two typical beam waists,
 $\mbox{w}_{\0}=1\,\mbox{mm}$ and $\mbox{w}_{\0}=2\,\mbox{mm}$, and for a wavelength value
 of $633\,\mbox{nm}$. The plot shows the wave number spread and the spatial laser width
 are complementar.}
\end{figure}

\newpage

\begin{figure}[hbp]
\hspace*{-2.5cm}
\includegraphics[width=19cm, height=22cm, angle=0]{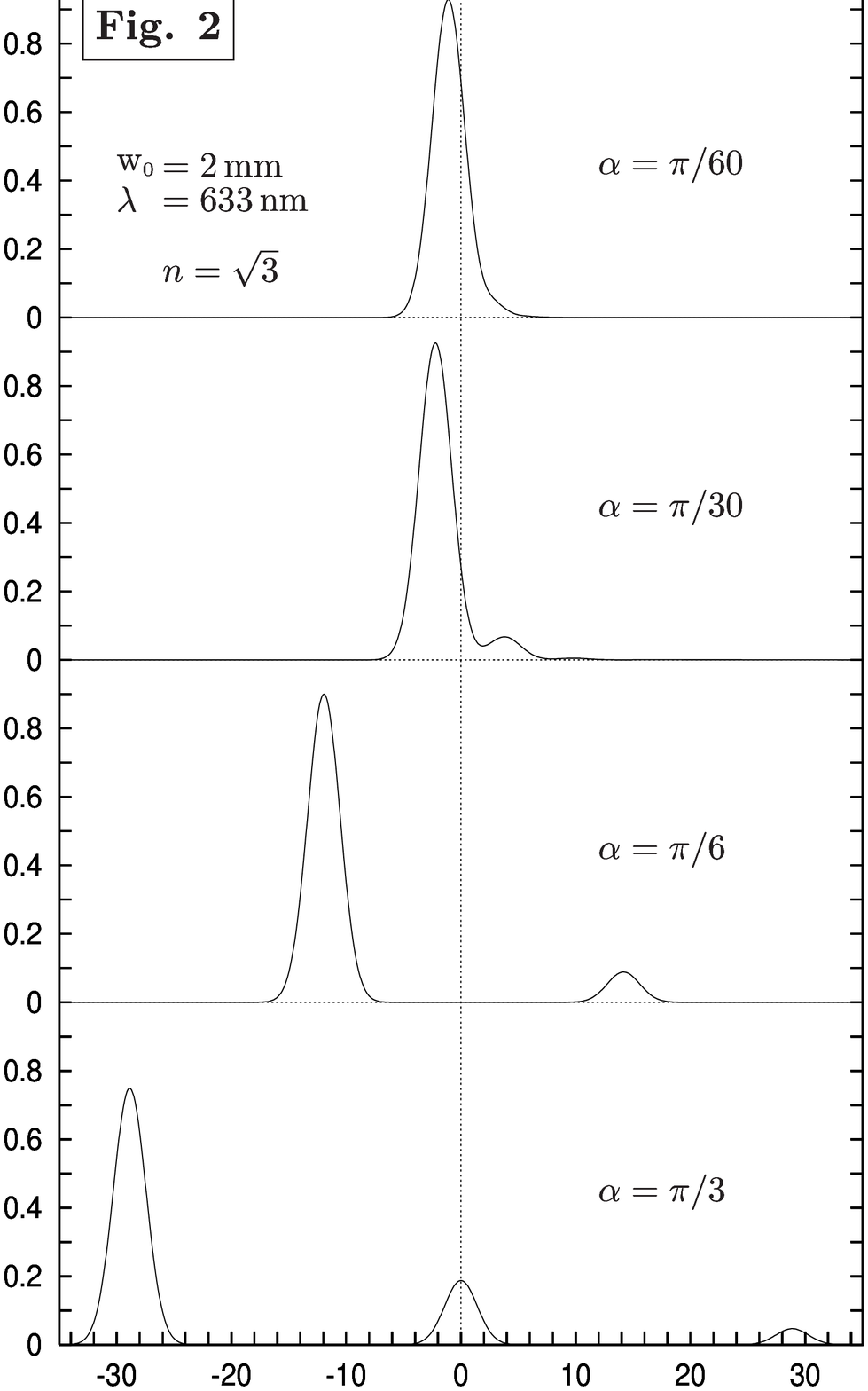}
\vspace*{-1.8cm}
 \caption{Numerical profiles of the modulus of the transmitted (normalized) electric
  field amplitude, for $x=0$ and $z=1\,\mbox{m}$, as a function of $y$. Four values of incident
  angles are shown. For increasing $\alpha$ the appearance of secondary peaks is seen. }
\end{figure}

\newpage

\begin{figure}[hbp]
\vspace*{-3cm}
 \hspace*{-2cm}
\includegraphics[width=19cm, height=28cm, angle=0]{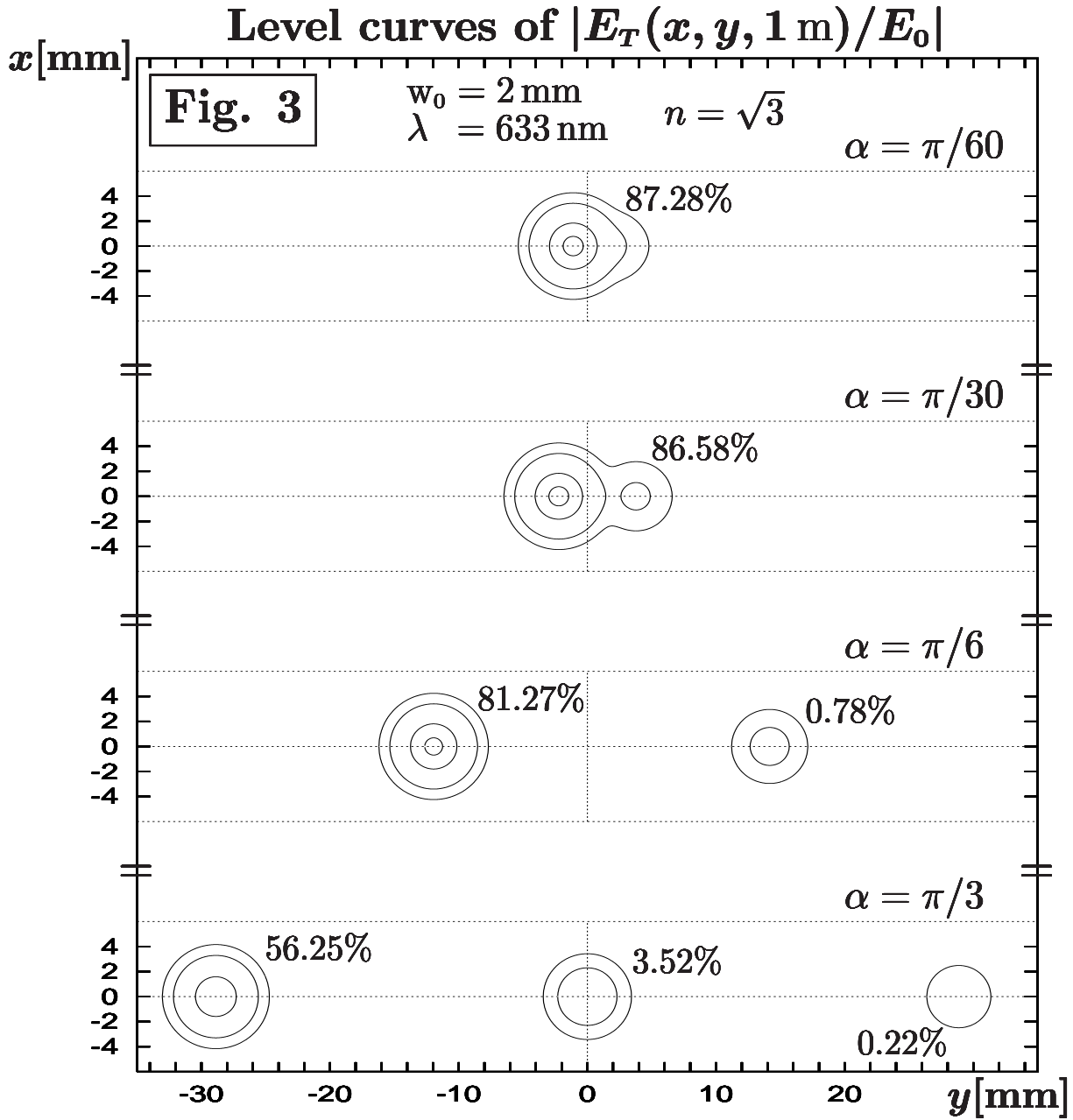}
\vspace*{-6.8cm}
 \caption{Planar view of the transmitted laser profiles displayed in Fig.\,2.
 The outer curve corresponds to an intensity density of $0.01$ compared to the
 intensity of the incident peak, the second curve, where it exists,
 corresponds to $0.05$, the third to $0.4$ and the fourth to $0.8$. For each beam,
 we also give the numerical value of the integrated relative power $P_{rel}$
 expressed in Eq.\,(\ref{prel}).}
\end{figure}

\end{document}